\documentclass[aps,pre,reprint,noeprint,showpacs,showkeys]{revtex4-1} 
\usepackage[utf8]{inputenc}
\usepackage[UKenglish]{babel}

\usepackage{lmodern}
\usepackage[T1]{fontenc}
\usepackage{microtype}
\usepackage{amsmath,amssymb,mathtools}

\usepackage{graphicx,grffile,textcomp}
\graphicspath{{figures/}}
\newlength{\figwidth}
\usepackage{url,hyperref}
\usepackage[usenames,dvipsnames]{xcolor}
\hypersetup{colorlinks=true, linkcolor=BrickRed, urlcolor=blue!50!black, citecolor=blue!50!black, unicode=true}
\hypersetup{bookmarksdepth=0} 

\usepackage{cleveref}
\crefrangelabelformat{equation}{(#3#1#4)$-$(#5#2#6)}

\renewcommand\epsilon{\varepsilon}
\renewcommand\phi{\varphi}
\renewcommand\theta{\vartheta}
\renewcommand\rho{\varrho}
\renewcommand\vec[1]{\textrm{\bfseries #1}}

\newcommand\diff{\mathrm{d}}
\newcommand\expect[1]{\left\langle{#1}\right\rangle}
\newcommand\e{\text{e}}
\renewcommand\i{\text{i}}

\newcommand\kB{\ensuremath{k_{\text{B}}}}
\mathchardef\mhyphen="2D
\newcommand\ellv{\ensuremath{\ell\mhyphen v}}
\newcommand\Hint{\ensuremath{\mathcal{H}_{\text{int}}}}

\newcommand\mathscr[1]{\scalebox{.9}{\ensuremath{\mathcal{\uppercase{#1}}}}}

\newcommand\bending{\mathscr{k}}

\begin{document}
\title{Enhanced wavelength-dependent surface tension of liquid--vapour interfaces}

\author{F. Höf{}ling}
\author{S. Dietrich}

\affiliation{
  Max-Planck-Institut für Intelligente Systeme, Heisenbergstraße 3,
  70569 Stuttgart, Germany, and \\
  IV. Institut für Theoretische Physik,
  Universität Stuttgart, Pfaffenwaldring 57, 70569 Stuttgart, Germany
}

\begin{abstract}
Due to the simultaneous presence of bulk-like and interfacial fluctuations the
understanding of the structure of liquid--vapour interfaces poses a
long-lasting and ongoing challenge for experiments, theory, and simulations.
We provide a new analysis of this topic by combining high-quality simulation
data for Lennard-Jones fluids with an unambiguous definition of the
wavenumber-dependent surface tension $\gamma(q)$ based on the two-point
correlation function of the fluid.
Upon raising the temperature, $\gamma(q)$ develops a maximum at short
wavelengths.
We compare these results with predictions from density functional theory.
Our analysis has repercussions for the interpretation of grazing-incidence
small-angle X-ray scattering (GISAXS) at liquid interfaces.
\end{abstract}


\maketitle

\paragraph*{Introduction.---}

\begin{figure*}
  \begin{center}
  \includegraphics[resolution=600]{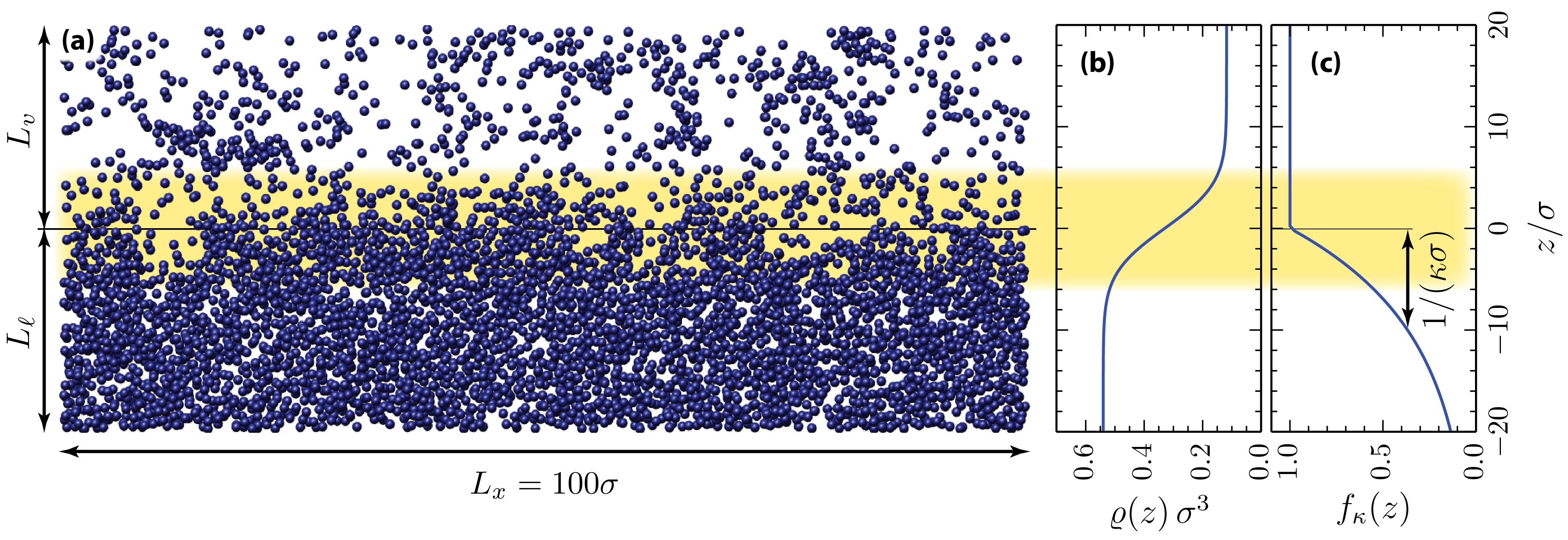}
  \end{center}
  \caption{Simulation snapshot of the \ellv{} interface for a LJ fluid at $T^* = 1.15$.
  A vertical slice of width 3.5$\sigma$ and height 40$\sigma$ is shown
  with the interfacial region highlighted (panel~a); $\sigma$ is the particle diameter.
  Panel~(b) displays the corresponding mean density profile
  $\rho(z)$ and panel~(c) the weight function $f_\kappa(z)$ used to compute the GISAXS
  intensity from \cref{eq:gid_master} for $\kappa\sigma=0.1$.
  }
  \label{fig:snapshot}
\end{figure*}

\noindent In a simple fluid, liquid ($\ell$) and vapour ($v$) phases can
coexist at temperatures between those of the triple point and the \ellv{}
critical point, $T_t  \leqslant T \leqslant T_c$.
A planar \ellv{} interface gives rise to a free energy $\gamma_0 A$ in excess
of the bulk free energy; $A$ is the interface area and $\gamma_0$ the
macroscopic surface tension between the two coexisting bulk phases.
The interface is spatially broadened by the simultaneous occurrence of
bulk-like fluctuations of the local number density and of thermally excited
capillary wave (CW)-like fluctuations~\cite{Evans:1979, Rowlinson:Capillarity,
Henderson_JR:1992} (\cref{fig:snapshot}a).
The free energy cost of the latter has been predicted to depend on their
wavenumber~$q$ and has been expressed in terms of a $q$-dependent
surface tension $\gamma(q)$ entering into an effective interface Hamiltonian
\Hint~\cite{Napiorkowski:1993, Parry:1994, Mecke:1999}, which allows one to
calculate the spatial structure of the interface.
Within density functional theory (DFT), a pronounced minimum of $\gamma(q)$ was
found as a consequence of the omnipresent van der Waals forces between the
molecules of the fluid~\cite{Mecke:1999}:
$\gamma(q\to 0) = \gamma_0 + a q^2 \ln(q b) + O\bigl(q^4\bigr)$ with
$a>0$~\cite{Napiorkowski:1993} and a length $b$.
The initial non-analytic decrease reflects the asymptotic decay $\sim r^{-6}$
of the pair potential.
The length $b$ is determined by compressibility effects in the interfacial
region and thus it is sensitive to microscopic details of the interaction; it
controls both the position $q_\text{min} = (b \sqrt{\e})^{-1}$ of the minimum
and its depth.
Evidence for such a minimum has been obtained experimentally from GISAXS on the
\ellv{} interface of water~\cite{Fradin:2000}, other molecular
fluids~\cite{Mora:2003}, and liquid gallium~\cite{Li:2004}.
The variation of $\gamma(q)$ signals that, depending on their wavelength,
interfacial fluctuations are either \emph{favoured} or \emph{suppressed} with
potentially far-reaching implications for nanoparticle
probes~\cite{Bickel:2014, Blanc:2013}, wetting films~\cite{MacDowell:2014}, and
droplet nucleation~\cite{Malijevsky:2012}.
Although liquid droplets are \emph{a priori} distinct from flat interfaces,
their surface tension varies with the radius~\cite{Henderson_JR:1992,
Block:2010, Das:2011} and one may expect a related non-analyticity therein for
long-ranged interactions~\cite{Bieker:1998, Blokhuis:2013}.


In simulations, a \emph{decrease} of $\gamma(q)$ for decreasing wavelength
$2\pi/q$ has been observed for polymeric liquids \cite{Mueller:2000,
Milchev:2002} and for colloidal suspensions~\cite{Vink:2005, Blokhuis:2008,
Blokhuis:2009}; an \emph{increase} has been found for
water~\cite{Sedlmeier:2009} and simple fluids~\cite{Chacon:2003}.
These studies are based on short-ranged interactions and yield monotonic
functions $\gamma(q)$ so that the aforementioned minimum could not be
corroborated by simulations so far.
New evidence for untruncated Lennard-Jones (LJ) fluids near $T_t$
suggests that there is indeed a shallow minimum~\cite{Chacon:2014}.
These simulation data have almost always been analysed in terms of the height
fluctuations of the local position $\hat z(\vec R)$ of an intrinsic interface
as considered also within DFT studies.
It requires interpreting a given configuration of particle positions
$\{ \vec r_j = (\vec R_j, z_j) \}_{j=1,\dots,N}$ in terms of a fluctuating
interface $\hat z(\vec R)$ with an associated intrinsic density profile, i.e.,
the microscopic number density at point $\vec r=(\vec R=(x,y), z)$ is assumed
to take the form
$\hat\rho(\vec R, z) = \hat \rho_\text{int}\boldsymbol (z-\hat z(\vec R)\boldsymbol )$.
The definition of $\hat z(\vec R)$ is not unique, leading to an ambiguity in
$\gamma(q)$~\cite{Parry:1994, Parry:2014, *Parry:2014a}.
Even the qualitative shape of $\gamma(q)$ depends on the choice of $z(\vec
R)$~\cite{Tarazona:2007, Tarazona:2012}: for a LJ fluid, the local version of
Gibbs' dividing surface (GDS) criterion yields a \emph{decreasing} $\gamma(q)$,
while $\gamma(q)$ \emph{increases} monotonically if a many-particle
definition is employed~\cite{Chacon:2003}.
Such kind of ambiguities have to drop out from expressions for
interface-related physical observables.
In particular, scattering data do not depend on various notions of an intrinsic
interface.


Here, we discuss $\gamma(q)$ in terms of the interface structure factor $H(q)$.
It allows one to extract $\gamma(q)$ from GISAXS data without additional
assumptions concerning the sample or choosing specific models for it.
Our simulation results enrich the physical picture of \ellv{} interfaces
considerably: Near the triple point, $\gamma(q;T\approx T_t)$ is almost flat
over a wide $q$-range.
Sweeping $T$ from $T_t$ to the critical point $T_c$, $\gamma(q;T)$ shows a
significant enhancement and develops a maximum at short wavelengths.
This behaviour can be rationalised in terms of a length scale which grows upon
approaching~$T_c$.

\paragraph*{Definition of $\gamma(q)$.---}

The structure of the \ellv{} interface separating the coexisting bulk phases is
characterised by the two-point density correlation function~\cite{Evans:1979}
$
  G(|\vec q|, z, z')=
    \!\int \! \diff^2 R \, \e^{-\i \vec q \cdot \vec R} \,
    \bigl[\expect{\hat\rho(\vec 0, z) \hat\rho(\vec R, z')} - \rho(z) \rho(z')\bigr]
$
in planar geometry; $\vec q=(q_x, q_y)$ is the lateral wavevector, and the mean
density profile $\rho(z)= \expect{\hat\rho(\vec R, z)}$ interpolates between
the number densities $\rho_\ell$ and $\rho_v$ of the coexisting phases
(\cref{fig:snapshot}b).
Although $G(q, z, z')$ is not directly accessible to experiments, information
on it can be obtained from GISAXS: the scattered intensity for a lateral
scattering vector~$\vec q$ is proportional to
[eq.~(2.68) in ref.~\citenum{Dietrich:1995}]
\begin{equation}
  I_\text{tot}(\vec q; \kappa) = \int \hspace{-2ex} \int \! \diff z \,\diff z' \, f_\kappa(z)^* \, f_\kappa(z') \, G(\vec q, z, z') \,,
  \label{eq:gid_master}
\end{equation}
omitting reflection and transmission coefficients.
The function $f_\kappa(z)$ depends on the scattering geometry, i.e., the angles
$\alpha_i$ and $\alpha_f$ of incidence and detection, respectively.
For X-rays entering from the vapour side, $f_\kappa(z)$ describes the decay of
the associated evanescent wave on the liquid side,
$f_\kappa(z<0) = \exp(-\kappa |z|)$ with the penetration depth
$1/\kappa(\alpha_i,\alpha_f)$ \cite{Dietrich:1984, Dietrich:1995}
(\cref{fig:snapshot}c).
Here, $z=0$ indicates the position of the GDS for $\rho(z)$.
Due to the low number density of the vapour, we take $f_\kappa(z>0)=1$; for
improvements see ref.~\cite{Dietrich:1995}.
For $\alpha_i$ and $\alpha_f$ close to the critical angle of total reflection,
the X-rays penetrate the sample deeply ($\kappa \to 0$) and the total scattered
intensity
$I_\text{tot}(q; \kappa)=I_\text{b}(q; \kappa) + I_\text{int}(q; \kappa)$
contains a sizeable contribution $I_\text{b}(q; \kappa)$ from bulk fluctuations
on top of the interfacial one, $I_\text{int}(q; \kappa)$.
A sufficiently large penetration depth $1/\kappa \gg \zeta$ is required in
order to obtain complete information on the interface of width $\zeta$; if the
interfacial region probed is too shallow, (rare) CWs with a large amplitude are
not captured by the X-rays.
The bulk scattering intensity
$I_\text{b}(q; \kappa) = I_\text{b}^{(\ell)}(q; \kappa) + I_\text{b}^{(v)}(q)$
is determined by $G(q,z,z')$ for $|z|, |z'| \to \infty$, i.e., by the familiar
structure factors $S_\alpha(k)$ (see below) of both phases $\alpha=\ell, v$
[c.f., \cref{eq:gid_bulk}];
in particular, one finds that the liquid side yields
$I_\text{b}^{(\ell)}(q; \kappa) = \rho_\ell S_\ell(q) /(2 \kappa) +
O\bigl(\kappa^0\bigr)$ as $\kappa \to 0$.
On the other hand, the interface part $I_\text{int}(q; \kappa)$ has to become
independent of the penetration depth as $\kappa \to 0$, which motivates and
facilitates the definition of an interface structure factor $H(q)$ as
$I_\text{int}(q; \kappa) = H(q) + O(\kappa)$, which is
equivalent to
\begin{equation}
  H(q) := \lim_{\kappa \to 0} \left[ I_\text{tot}(q; \kappa) - I_\text{b}(q; \kappa) \right] .
  \label{eq:H_q}
\end{equation}
Upon construction, $H(q)$ is specified by $G(q, z, z')$ alone. It summarises
the modifications of the local density distribution caused by the presence of
the interface.

Next, we consider $H(q\to 0)$ within classical CW theory, which---in the
absence of an external potential---predicts a divergence of
$
  G(q\to 0,z,z') \simeq \kB T \rho'(z) \rho'(z') / \bigl(\gamma_0 q^2\bigr)
$
\cite{Buff:1965, Wertheim:1976, Weeks:1977, Evans:1979}.
We insert this into \cref{eq:gid_master,eq:H_q} and use that
$I_\text{tot}(q; \kappa) = I_\text{b}(q; \kappa) + O\bigl(\kappa^0\bigr)$
as $\kappa\to 0$ for $q$ fixed and that $I_\text{b}(q; \kappa)$ is bounded in
$q$ for $\kappa$ fixed.
Since $q^2 I_\text{int}(q; \kappa)$ is regular for $\kappa, q \to 0$
one finds asymptotically \cite{GID:2014}
\begin{equation}
  H(q \to 0) \simeq \kB T \, (\Delta \rho)^2 \left[\gamma_0 \, q^2 \right]^{-1} ,
  \label{eq:H_CW}
\end{equation}
where $\Delta\rho :=\rho_\ell - \rho_v$.
For $\kappa \zeta \ll 1$ but fixed, $H(q)$ dominates the scattering intensity
as $q\to 0$: $I_\text{tot}(q; \kappa) = H(q) + O\bigl(q^0\bigr)$.
Thus in principle, $\gamma_0$ can be inferred from $I_\text{tot}(q \to 0;
\kappa)$ directly.
We generalise \cref{eq:H_CW} to arbitrary $q$, such that
$\gamma(q\to 0)=\gamma_0$, by \emph{defining} a
$q$-dependent surface tension
\begin{equation}
  \gamma(q) := \kB T (\Delta \rho)^2 \bigl[q^2 H(q) \bigr]^{-1} \,.
  \label{eq:gamma_q_def}
\end{equation}
Therefore, $H(q)$ and $\gamma(q)$ are equivalent descriptions of the interface
structure, both determined by $G(q, z, z')$.

\paragraph*{Bulk contribution.---}

The bulk contribution $I_\text{b}(q; \kappa)$ follows from the corresponding
expression for $G_\text{b}(q, z, z')$ entering \cref{eq:gid_master}.
The bulk reference system is formed by two independent, macroscopically large
half-spaces $z< 0$ and $z > 0$ filled with the coexisting liquid and vapour
phase, respectively; the bulk structures are taken to be unperturbed by the
sharp boundary at $z=0$.
Thus $G_\text{b}(q, z, z')$ equals the bulk functions $G_\alpha(q,|z-z'|)$ if
$z$ and $z'$ are on the same side of the interface ($zz' > 0$) and
$G_\text{b}(q,z,z')=0$ otherwise; $G_\alpha(q,|z-z'|)$ is related uniquely to
$
  S_\alpha(|\vec k|) = \rho_\alpha^{-1} \int_{-\infty}^\infty \e^{-\i k_z u} \,
  G_\alpha(|\vec q|,u) \,\diff u + \rho_\alpha (2\pi)^3 \delta(\vec k)
$
for $\vec k = (\vec q, k_z)$.
All correlations between particles located on opposite sides of the interface
are contained exclusively in $H(q)$, and $I_\text{b}(q; \kappa)$ is the sum of
the scattered intensities from the two half-spaces.
For simulations, the finiteness of the scattering volume of height $L_\alpha$
for either phase (\cref{fig:snapshot}b) must also be taken into account,
leading to~\cite{GID:2014}
\begin{multline}
  I_\text{b}^{(\alpha)}(q; \kappa) =
    \rho_\alpha \int_{-\infty}^\infty \! \frac{\diff k_z}{2 \pi} \,
    \frac{S_\alpha\bigl(\sqrt{q^2 + k_z^2}\bigr)}{\kappa^2 + k_z^2} \\
  \times \left[1 + \e^{-2\kappa L_\alpha} - 2 \e^{-\kappa L_\alpha} \cos(k_z L_\alpha) \right]
  \label{eq:gid_bulk}
\end{multline}
up to a term $\propto \delta(\vec q)$; for $\alpha=v$, $\kappa = 0$.
Concerning the interpretation of future GISAXS data, the second line is
replaced by unity ($L_\alpha\to \infty$).

It is instructive to inspect the asymptotics of $I_\text{b}^{(\alpha)}$ with
respect to the non-commuting limits
$\kappa \to 0, L_\alpha \to \infty$~\cite{GID:2014}:
\begin{equation*}
  I_\text{b}^{(\alpha)}(q; \kappa) \simeq
  \frac{1-\e^{-2\kappa L_\alpha}}{2\kappa} \, \rho_\alpha S_\alpha(q)
    + \left(1+\e^{-2\kappa L_\alpha}\right) \, \rho_\alpha \mathcal{I}_0^{(\alpha)}(q)
\end{equation*}
up to terms $O(\kappa)$, $\e^{-\kappa L_\alpha} O\bigl(L_\alpha^{-1}\bigr)$.
Taking $L_\ell \to \infty$ first, the first term on the l.h.s.\ describes the
singularity of the bulk scattering from the liquid side:
$I_\text{b}^{(\ell)}(q; \kappa) \simeq \rho_\ell S_\ell(q)/(2\kappa)$ as
$\kappa \to 0$.
If $\kappa \to 0$ first (as for $\alpha=v$),
$I_\text{b}^{(\alpha)}(q; \kappa) \simeq \rho_\alpha L_\alpha\, S_\alpha(q)$
as $L_\alpha \to \infty$.
The next order $\sim (\kappa L)^0$ contains
$
  \mathcal{I}_0^{(\alpha)}(q)
  := \int \! \diff k_z \,
      \bigl[S_\alpha\bigl(\sqrt{q^2 + k_z^2}\bigr) - S_\alpha(q) \bigr] / \bigl(2\pi k_z^{-2} \bigr) ,
$
which are bounded functions of $q$.
According to \cref{eq:H_q}, these terms contribute to $H(q)$, and it is crucial
that the full expression for $I_\text{b}(q; \kappa)$ is used therein.
In particular, adopting a definition for $G_\text{b}(q, z, z')$ and thus
$I_\text{b}(q; \kappa)$ different from \cref{eq:gid_bulk} affects $\gamma(q)$.
Whereas the leading terms of $I_\text{b}(q; \kappa)$ and
$I_\text{tot}(q; \kappa)$ must coincide, there is no such constraint for the
next order.
For example, including bulk-like correlations due to the presence of the
interface, $G_\text{b}(q,z,z') \neq 0$ for $zz' < 0$, modifies $\gamma(q)$ by a
term $O\bigl(q^2\bigr)$ \cite{Parry:2014, *Parry:2014a}.
Previous interpretations of GISAXS experiments~\cite{Fradin:2000, Mora:2003}
approximated
$I_\text{b}^{(\ell)}(q; \kappa) \approx  \rho_\ell S_\ell(0) / (2\kappa)$,
which for $\gamma(q)$ implies an extra factor of
$
  1 + q^2 (\gamma_0 / \kB T) (\Delta \rho)^{-2} \rho_\ell \mathcal{I}_0^{(\ell)}(0)
  + O\bigl(q^4\bigr) \,;
$
it reduces $\gamma(q)$ if $\mathcal{I}_0^{(\ell)}(0) <0$.

\paragraph*{Simulation method.---}

\newbox{\widetildeH}
\savebox\widetildeH{$\widetilde H$}

\begin{figure}
  \includegraphics[width=\figwidth]{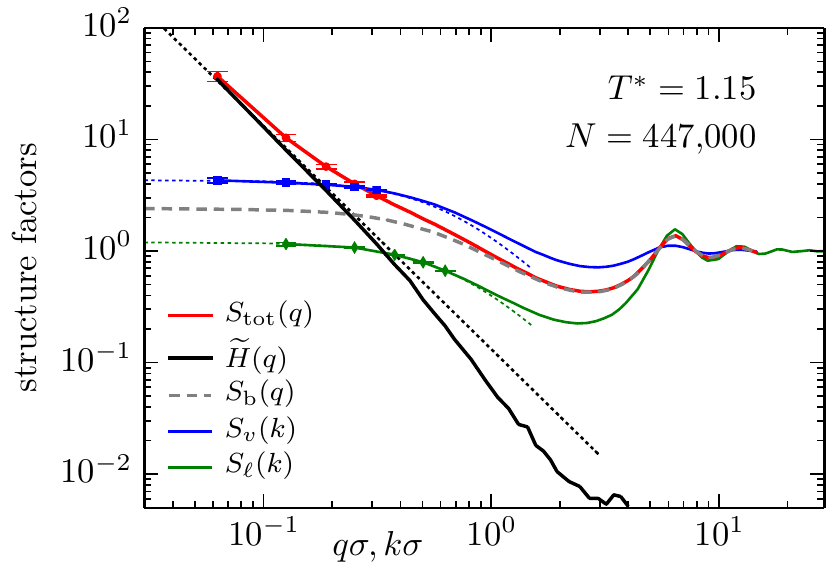}
  \caption{%
  Decomposition of $S_\text{tot}(q)=(\mathcal{A} / N) I_\text{tot}(q; \kappa \to 0)$
  for the inhomogeneous system ($L_\ell=50\sigma, L_v = 3 L_\ell$) into the interface structure
  factor \usebox{\widetildeH}$(q)=(2\mathcal{A}/N) H(q)$ and the bulk contribution $S_\text{b}(q)$.
  The latter follows from the structure factors $S_\ell(k)$ and $S_v(k)$
  of the coexisting bulk phases [\cref{eq:gid_bulk}]; the bulk data are
  extrapolated to $k=0$ by an Ornstein--Zernike fit (green and blue dotted
  lines).
  Solid lines connect actual data points.
  The black dotted line, which is approached by \usebox{\widetildeH}$(q\to 0)$,
  indicates the prediction of classical CW theory  [\cref{eq:H_CW}].
  }
  \label{fig:ssf}
\end{figure}

\begin{figure*}
  \hfill \includegraphics{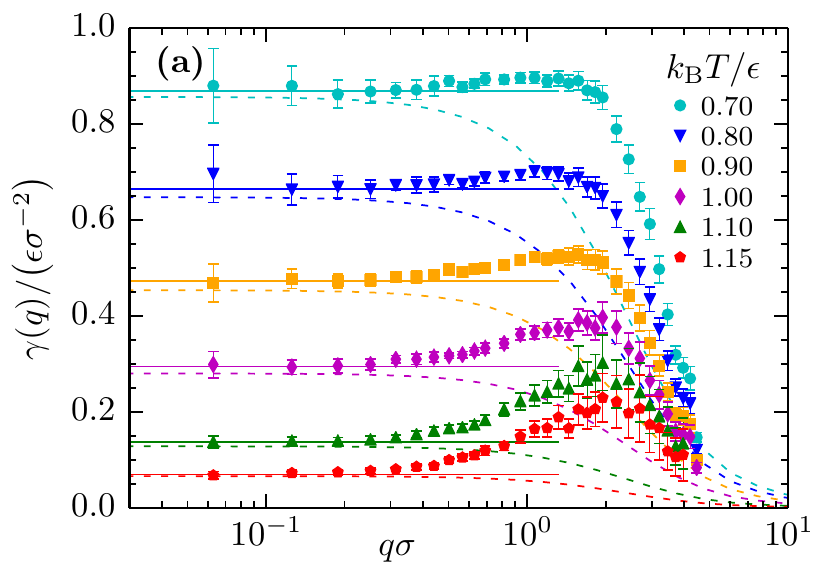} \hfill
  \includegraphics{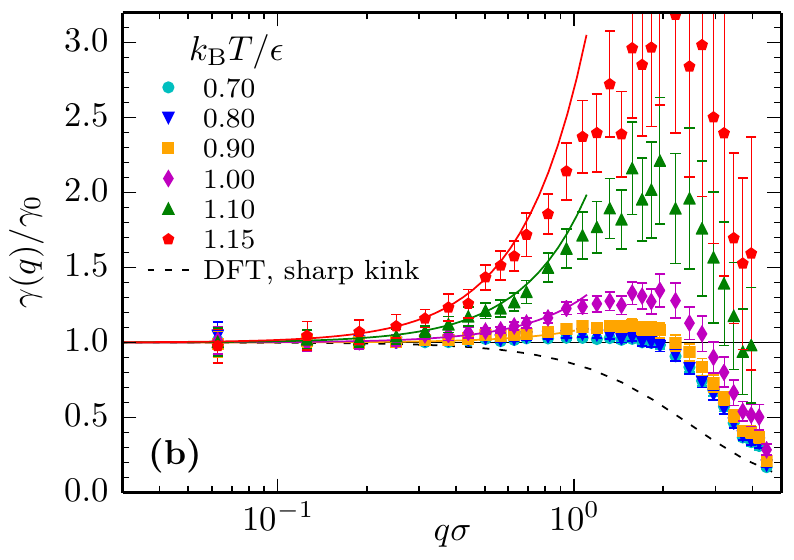} \hfill\hbox{}
  \caption{Wavenumber-dependent surface tension $\gamma(q)$ of LJ
  fluids with truncated interaction range ($r_c=3.5\sigma$) as extracted from
  \usebox{\widetildeH}$(q)$ according to \cref{eq:gamma_q_def}.
  \enspace (a) Solid horizontal lines indicate the macroscopic surface tensions
  $\gamma_0(T)$ determined via the mechanical route.
  Dashed lines represent the simulation-enhanced DFT prediction [see
  the main text and \cref{eq:gamma_q_DFT}]; $\gamma_0(T)$ from MD simulations
  and from theory differ slightly at low $T$.
  \enspace (b) The same data for $\gamma(q)$ normalised by $\gamma_0(T)$; full
  lines are fits of a quadratic increase [\cref{eq:gamma_quadratic}].
  The black dashed line shows the DFT prediction within the sharp-kink
  approximation~\cite{Napiorkowski:1993}.
  }
  \label{fig:gamma_q_rc3.5}
\end{figure*}

For LJ fluids we have performed molecular dynamics (MD) simulations of a flat
liquid slab coexisting with its vapour phase. The LJ pair potential
$V(r)=4\epsilon\left[(r/\sigma)^{-12} - (r/\sigma)^{-6}\right]$ is truncated
and shifted to zero at $r_c=3.5\sigma$ (not $2.5\sigma$ as usual).
The investigated dimensionless temperatures range from
$T^*:=\kB T/\epsilon=0.70$ slightly above $T_t^*\approx 0.68$
\cite{Hansen:1969} up to
$T^*=1.15\approx 0.94 \, T_c^*$.
We have estimated $T_c^*\approx 1.22$ from the linear extrapolation to 0 of
$(\Delta \rho)^{1/\beta}$ as function of $T$, using $\beta=0.327$
\cite{Pelissetto:2002, Watanabe:2012}.
Since we are interested, \emph{inter alia}, in small $q$, large simulation
boxes containing several 10\textsuperscript{5} particles are needed.
Powerful means for such demanding simulations are provided by the highly
parallel architecture of modern graphics processors (GPUs).
We have used the software \textsl{HAL's MD package} (version
0.3)~\cite{Glassy_GPU:2011, *HALMD}, which delivers accurate and efficient,
GPU-accelerated MD simulations.\footnote{%
  Employing one Tesla K20Xm GPU (NVIDIA Corp.), a simulation of 447,000
  particles at $T^*=1.15$ over 10\textsuperscript{7} steps took about 28\,h,
  including the computation of $I_\text{tot}(q; \kappa)$.
}
The geometry of the simulation box is given by the total height
$L_z=L_\ell + L_v$ and the cross-sectional area~$\mathcal{A}=L_x L_y$; the
total number of particles $N$ fixes the ratio $L_\ell/L_v$ via the GDS
criterion $N / \mathcal{A} =\rho_\ell L_\ell + \rho_v L_v$.
Two parallel, on average flat \ellv{} interfaces of total area $A=2
\mathcal{A}$ have been prepared by bringing separately equilibrated, coexisting
bulk phases into spatial contact (with periodic boundary conditions in all
directions).
We have used liquid slabs of width $L_\ell=25\sigma$ ($T^* \leqslant 1.0$) or
$50\sigma$ ($T^* \geqslant 1.1$), $L_x=L_y=100\sigma$, and $L_z=125\sigma$ or
$200\sigma$.
In the next step, the thermally excited CWs have been equilibrated by MD runs
lasting $3{-}10 {\times} 10^3 \tau$ with $\tau=(m\sigma^2/\epsilon)^{1/2}$ and
particle mass~$m$.
The production runs have covered at least $15 {\times} 10^3 \tau$, and averages
over 10$-$30 independent runs have been taken.
In order to minimise disk access we have used the H5MD file
format~\cite{H5MD:2014} and have evaluated all relevant observables \emph{in
situ}, e.g.,
$I_\text{tot}(|\vec q|; \kappa) = \expect{|\hat\rho_\kappa(\vec q)|^2} / \mathcal{A}$
where
$\hat\rho_\kappa(\vec q) = \sum_{j=1}^N f_\kappa(z_j)\, \e^{\i \vec q \cdot \vec R_j}$
is computed after each interval~$\tau$.
Further, we introduce the dimensionless quantities\footnote{%
  We emphasise that $S_\text{b}(q) \neq \rho_\ell L_\ell S_\ell(q) + \rho_v L_v S_v(q)$.
}
$S_\text{tot}(q) := (\mathcal{A} / N) I_\text{tot}(q; \kappa\to 0)$,
$S_\text{b}(q) := (\mathcal{A} / N) I_\text{b}(q; \kappa\to 0)$,
and $\widetilde H(q) := S_\text{tot}(q) - S_\text{b}(q) = (2\mathcal{A}/N) H(q)$.

\Cref{fig:ssf} illustrates the determination of $\widetilde H(q)$ for
$T^*=1.15$.
The total structure factor $S_\text{tot}(q)$ displays a steep increase for
$2 \gtrsim q\sigma \to 0$, and at $q=2\pi/L_x$ it exceeds the height of the
first peak of the liquid structure by a factor of ca.~27.
Nevertheless, it barely displays the CW asymptote $\sim q^{-2}$
[\cref{eq:H_CW}].
The bulk contribution $S_\text{b}(q)$ is computed via \cref{eq:gid_bulk} with
$S_\ell(k)$ and $S_v(k)$ obtained from separate simulations of homogeneous
systems at $\rho_\ell\sigma^3=0.540$ and $\rho_v \sigma^3=0.118$.
The functions $S_\alpha(k)$ are extrapolated to $k= 0$ using the
Ornstein--Zernike form
$S_\alpha(k) \simeq \rho_\alpha \kB T \chi_\alpha / \bigl[1+(k\xi_\alpha)^2\bigr]$
by fitting the bulk compressibilities~$\chi_\alpha$ and the correlation
lengths~$\xi_\alpha$.
The highly compressible vapour phase contributes about 70\% to
$S_\text{b}(q\to 0)$ and must not be ignored; it still adds 22\% at
$T^*=0.70$.\footnote{%
  Note that the vapour contribution has been neglected in refs.~\cite{Blokhuis:2008,
  Blokhuis:2009, Sedlmeier:2009}, which at small $q$ amounts to modifying
  $\gamma(q)$ by a term $O(q^2)$.
}
For $q\sigma \gtrsim 3$, the bulk part $S_\text{b}(q)$ resembles
$S_\text{tot}(q)$ with a deviation of less than 1\%.
The signal of $\widetilde H(q)$ is visible over more than 3 decades and clearly
approaches and follows the CW divergence as $q\to 0$ [\cref{eq:H_CW},
\cref{fig:ssf}].
The only parameter left unspecified is $\gamma_0$, which we have determined
independently via the mechanical route~\cite{Evans:1979}.
With this, the whole procedure does not involve any adjustable parameter.

\paragraph*{Results for $\gamma(q)$.---}

The $q$-dependent surface tension follows from \cref{eq:gamma_q_def}.
We have generated high-quality data for $\gamma(q;T)$ for temperatures $T$
covering almost the whole two-phase coexistence region and for wavenumbers~$q$
spanning 2 decades (\cref{fig:gamma_q_rc3.5}a).
Upon raising $T^*$ from 0.70 to 1.15, the macroscopic surface tension
$\gamma_0(T)$ decreases by a factor of 12.
The concomitant reduction of $\gamma(q)$ is, however, not uniform for all~$q$:
it is only about a factor of~4 for $q\sigma \approx 2$, leading to an
\emph{enhancement} of $\gamma(q)$ relative to $\gamma_0$.
The strong $T$-dependence can be accounted for by studying
$\gamma(q) / \gamma_0$ (\cref{fig:gamma_q_rc3.5}b).
At $T^*=0.70$, i.e., close to the triple point, $\gamma(q)/\gamma_0\approx 1$
for $q\sigma \lesssim 2$ and $\gamma(q)$ is nearly independent of~$q$.
As $T$ is raised, our data for $\gamma(q)$ change from an almost flat to an
initially increasing function.
For $q\sigma \gtrsim 2$, the surface tension is found to again decrease as a
function of~$q$.
Thus on route towards $T_c$, $\gamma(q)$ develops a \emph{maximum} at
$q_\text{max}(T)$, which shifts from $q_\text{max}\sigma \approx 1.0$ to $2.1$
as $T^*$ increases from 0.8 to 1.15.
The maximum value is enhanced over $\gamma_0$ by a factor of ca.~3 for
$T^*=1.15$.
The function  $\gamma(q)$ is analytic in $q^2$ due to $r_c < \infty$ and its
initial increase can be described by an emerging length~$\ell(T)$:
\begin{equation}
  \gamma(q;T)/\gamma_0(T) = 1 + \left[q\ell(T)\right]^2 + O\bigl(q^4\bigr) \,,
  \qquad q \to 0 \,.
  \label{eq:gamma_quadratic}
\end{equation}
The coefficient $\bending=\gamma_0 \ell^2$ has often been interpreted as a
bending rigidity of the interface; it has been shown within DFT for \Hint{}
that $\bending$ summarises compressibility effects of the coexisting
phases~\cite{Mecke:1999}.
Along these lines, the length $\ell$ may be viewed as an effective cutoff $q <
2\pi/\ell$ of the CW spectrum, which reduces the interface
roughness.~\cite{Tarazona:2012}.
For $T^*\geqslant 0.8$, we find $\ell(T)$ to be comparable to the bulk
correlation lengths $\xi_\ell$ and $\xi_v$.
The latter diverge as $\xi_\alpha \sim |t|^{-\nu}$ with
$\xi_\ell/\xi_v \to 1$, $\nu\approx0.630$, $t:=(T-T_c)/T_c \to 0$
\cite{Pelissetto:2002, Watanabe:2012},
and $\gamma_0 \sim |t|^{2\nu}$ \cite{Parry:1992, Watanabe:2012}. 
For $T^* \leq 1.15$, we find $\xi_\alpha \lesssim 1.5\sigma$ so that one cannot
assign a power law to these data.
If nonetheless one anticipates the same scaling for $\ell(T)$ and
$\ell/\xi_\alpha \to 1$, this implies
$\bending(T\to T_c) = (4\pi \omega)^{-1} k_B T_c$,
where $\omega\approx 0.87$ is a universal number~\cite{Hasenbusch:1993,
Das:2011}. 
Our data corroborate this scaling of $\bending$; however, it is at variance
with earlier predictions~\cite{Blokhuis:2008, Blokhuis:2009}.

\begin{figure}
  \includegraphics[width=\figwidth]{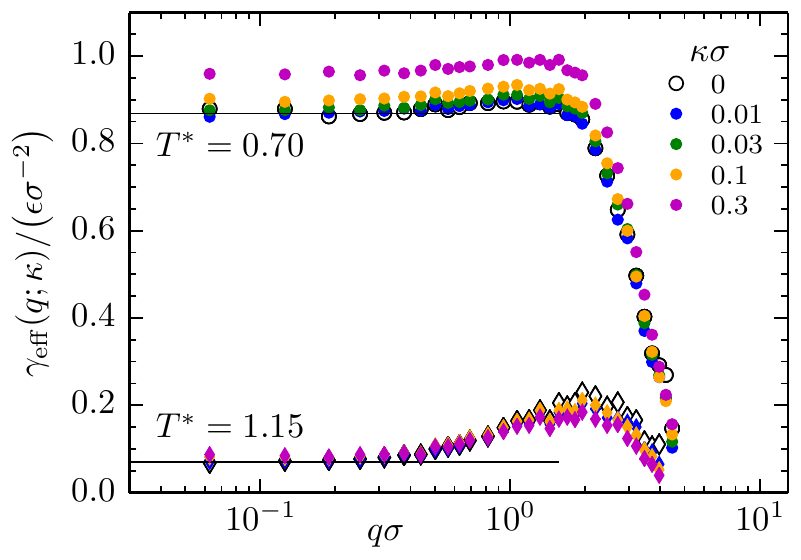}
  \caption{Dependence of $\gamma_\text{eff}(q; \kappa)$ on the penetration depth
  $1/\kappa$ at two temperatures.
  These results are based on the GISAXS intensity $I_\text{tot}(q; \kappa)$
  obtained on the fly from the MD simulations.
  The data for $\kappa = 0$ (open symbols) are reproduced from \cref{fig:gamma_q_rc3.5}.
  }
  \label{fig:gid_gamma_q_rc3.5}
\end{figure}

A pressing issue is the absence (in simulations) of the local minimum in
$\gamma(q)$ observed experimentally~\cite{Fradin:2000, Mora:2003, Li:2004} and
predicted theoretically~\cite{Mecke:1999} for fluids governed by \emph{bona
fide} dispersion forces decaying $\sim r^{-6}$.
Any interaction cutoff $r_c$ converts the corresponding non-analytic
contribution $q^2 \ln(qb)$ in $\gamma(q)$ into a term $-q^2 \ln(r_c/b)$, which
competes with the aforementioned term~$(q\ell)^2$.
However, $\gamma(q)$ is expected to not change qualitatively once $r_c$ is
sufficiently large ($q \gtrsim 2\pi/r_c$).
From recent simulations of LJ fluids, a shallow minimum near
$q_\text{min} \approx 0.24\sigma$ ($b\approx 2.5\sigma$) has indeed been
identified within a suitable extrapolation scheme for
$r_c \to \infty$~\cite{Chacon:2014}.
This is not accessible from our data.

\paragraph*{Simulations of GISAXS.---}

Within the MD simulations, we have also determined $I_\text{tot}(q; \kappa)$
for a range of penetration depths~$1/\kappa$ as it would be observed in GISAXS
experiments.
To this end, we define $\gamma_\text{eff}(q; \kappa)$ analogously to
\cref{eq:gamma_q_def}, therein replacing $H(q)$ by $I_\text{int}(q; \kappa)$.
It is crucial that both the finite-size effects and $\kappa>0$ are properly
included in $I_\text{b}(q; \kappa)$~[\cref{eq:gid_bulk}].
The data in \cref{fig:gid_gamma_q_rc3.5} support our mathematical finding that
$\gamma_\text{eff}(q; \kappa \to 0) = \gamma(q)$.
The increase of $\gamma_\text{eff}(q \to 0; \kappa)$ upon increasing $\kappa$
agrees quantitatively with the difference
$I_\text{int}(q; \kappa) - H(q) = O(\kappa\zeta)$.
The data for the present LJ fluids indicate that $\kappa \sigma \lesssim 0.03$
is small enough to yield a reliable approximation of $\gamma(q)$.

\paragraph*{Simulation-enhanced DFT.---}

For comparison with DFT predictions based on an effective interface Hamiltonian
\Hint~\cite{Mecke:1999}, we start from the systematic expansion of $\gamma(q)$
in terms of $q$ but neglect the Gaussian curvature contribution to the
intrinsic density profile (i.e., $C_H=0$ in the notation of
ref.~\cite{Mecke:1999}) because the corresponding terms in $\gamma(q)$ are not
amenable to a quantitative evaluation within our simulations.
The DFT expression left involves only $\rho(z)$ and the attractive part
$w(r) < 0$ of the pair potential:
\begin{equation}
  \hat\gamma(q) = -\frac{\pi}{2}
    \! \int \hspace{-2ex} \int_{-L_\ell}^{L_v} \hspace{-1em} \diff z \, \diff z' \,
    \rho'(z) \rho'(z')
    \! \int_0^{r_c} \hspace{-1em} \diff R \,R^3 \mathcal{J}(q R) \,w(r)
  \label{eq:gamma_q_DFT}
\end{equation}
up to terms $O\bigl(q^4, C_H q^2 \bigr)$, where
$\mathcal{J}(x) := 4[1-J_0(x)]/x^2$ with $J_0$ as a Bessel function of the
first kind and $r := \sqrt{R^2+(z_1-z_2)^2}$.
The limit $q\to 0$ of \cref{eq:gamma_q_DFT} corresponds to the
Triezenberg--Zwanzig result for $\gamma_0$~\cite{Triezenberg:1972,
Henderson_JR:1992} within the random phase approximation for the direct
correlation function,
$c(\vec r_1, \vec r_2)\approx -w(|\vec r_1 - \vec r_2|) / \kB T$.
We have evaluated \cref{eq:gamma_q_DFT} with $\rho(z)$ from the simulations
with $w(r \geqslant r_0) = V(r) - V(r_c)$ and $w(r < r_0) = V(r_0)- V(r_c)$;
$r_0 = 2^{1/6}\sigma$.
Over the whole $T$-range, we find good agreement between the prediction
$\hat\gamma(q=0)$ and the data for $\gamma_0(T)$ (\cref{fig:gamma_q_rc3.5}a).
Concerning the $q$-dependence, however, it turns out
(\cref{fig:gamma_q_rc3.5}b, ``DFT'') that 
$\hat\gamma(q;T)/\hat\gamma_0(T)$ depends only slightly on $T$ and coincides
nearly with the $T$-independent expression from the sharp-kink approximation,
which assumes a step-like variation of $\rho(z)$~\cite{Napiorkowski:1993}.
\Cref{eq:gamma_q_DFT} misses the observed change of the shape of $\gamma(q;T)$
upon varying $T$ and, moreover, yields $\bending/\gamma_0 \approx -0.16 < 0$ in
stark contrast to $\bending(T) > 0$ for all $T$ found above.
Note that the neglected term $O\bigl(C_H q^2\bigr)$, present in
DFT~\cite{Mecke:1999}, could render $\bending>0$.

Recently, Parry \emph{et al.}~\cite{Parry:2014, *Parry:2014a} have revisited the
\ellv-interface within an analytically tractable toy model and compared
various definitions of $\gamma(q)$.
The approach taken here is motivated by GISAXS experiments and takes the unique
interface structure factor $H(q)$ [obtained from $S_\text{tot}(q)$] as input
for~$\gamma(q)$.
Theoretical CW descriptions, on the other hand, employ a certain
$\Hint[\hat z(\vec R);\hat\gamma(q)]$,
from which one calculates the fluctuation spectrum $H_\text{CW}(q)$ of an
idealised sharp surface $\hat z(\vec R)$.
Within the toy model~\cite{Parry:2014, *Parry:2014a}, specifying
$\hat z(\vec R)$ by a local GDS or any other ``crossing criterion'' does not
generate the expected $H_\text{CW}(q)$.
Moreover for any choice of $\hat z(\vec R)$, the corresponding \Hint{} cannot
simply be related to $\gamma(q)$ as given by $H(q)$.
The latter summarises the experimentally accessible density fluctuations caused
by the presence of the interface, including non-CW (``bulk-like'') terms which
cannot be derived from any $\Hint$.\footnote{%
  The non-CW terms in $H(q)$ must not be confused with $S_\text{b}(q)$, the
  latter describing bulk fluctuations in the absence of an interface.
}
This reflects the fact that the representation of a \ellv{} interface by a
two-dimensional (2D) manifold renders only an incomplete picture of
this inherently 3D object.
Within our approach, $\gamma(q)$ provides a good characterisation of the 3D
structure of the interface as revealed by scattering experiments.

\paragraph*{Conclusions.---}

We have determined the q-dependent surface tension $\gamma(q;T)$ of a \ellv{}
interface covering the full two-phase coexistence region.
Our analysis is based on the interface structure factor $H(q)$ and resembles
the approach to interpret GISAXS data, thereby avoiding the ambiguous
identification of an instantaneous, fluctuating local interface position.
The decomposition of fluctuations into bulk and interfacial ones rests on the
distinct scaling behaviour of these two contributions upon increasing the
penetration depth of scattering waves (in experiments) or the system size (in
simulations).
As a complication, we found that the (non-singular) leading correction to the
bulk scattering intensity contributes to $H(q)$.
Our analysis provides a fresh view on GISAXS data for \ellv{}
interfaces~\cite{Fradin:2000, Mora:2003, Li:2004} as we do not require any
specific model for the bulk scattering; such models can render $\gamma(q)$
meaningless~\cite{Paulus:2008}.
Instead, the bulk contribution is determined via \cref{eq:gid_bulk} from the
familiar bulk structure factors, to be obtained from independent experiments or
simulations.
The surface tension $\gamma(q)$ discussed here is fully specified by the
two-point density correlation function $G(q,z,z')$ of the inhomogeneous system.
However, in general it differs from $\hat\gamma(q)$ entering \Hint{} and
describing the CW spectrum~\cite{Parry:2014, *Parry:2014a}.

Our extensive simulations reveal a pronounced $T$-dependence of the
shape of $\gamma(q;T)$ as a function of~$q$, which upon increasing $T$, at a
certain value~$q$, changes from an almost flat to a convex curve .
Our data indicate a positive ``bending'' coefficient $\bending=\gamma_0\ell^2$
[\cref{eq:gamma_quadratic}] for all $T$, facilitating its interpretation as the
square of an emerging length~$\ell(T)$.
For the simulated LJ fluids, we find a maximum of $\gamma(q;T)$ at wavelengths
$2\pi/q_\text{max}$ between $3\sigma$ and $6\sigma$, which builds up gradually
with an up to threefold \emph{enhancement} of $\gamma(q_\text{max};T)$ over
$\gamma_0(T)$ at $T\approx 0.94\,T_c$.
At elevated~$T$, measurements of $\gamma(q;T)$ using GISAXS are challenging due
to the significant vapour density.
However, \ellv{}-like interfaces in suitable colloidal suspensions can be
studied well~\cite{Aarts:2004} even close to criticality~\cite{Royall:2007},
which allows one to analyse $\gamma(q;T)$ along the lines presented here.

\acknowledgments

We thank R.~Evans, K.~Mecke, A.~Parry, M.~Oettel, and P.~Tarazona for
stimulating discussions and P.~Colberg for many valuable contributions to \textsl{HAL's MD
package.}
Some of the data were produced with the supercomputer \textsl{Hydra}
of the Max Planck society.

\bibliography{capillary}

\end{document}